\documentclass{aa}
\usepackage{graphicx}
\usepackage{txfonts}

\begin{document}
\title{The end of the white dwarf cooling sequence in M~67\thanks{
Based on data acquired using the Large Binocular Telescope (LBT) at
Mt. Graham, Arizona, under the Commissioning of the Large Binocular
Blue Camera.  The LBT is an international collaboration among
institutions in the United States, Italy and Germany. LBT Corporation
partners are: The University of Arizona on behalf of the Arizona
university system; Istituto Nazionale di Astrofisica, Italy; LBT
Beteiligungsgesellschaft, Germany, representing the Max-Planck
Society, the Astrophysical Institute Potsdam, and Heidelberg
University; The Ohio State University, and The Research Corporation,
on behalf of The University of Notre Dame, University of Minnesota and
University of Virginia; and on observations obtained at the
Canada-France-Hawaii Telescope (CFHT) which is operated by the
National Research Council of Canada, the Institut National des
Sciences de l'Univers of the Centre National de la Recherche
Scientifique of France, and the University of Hawaii.}}

\author{Bellini, A.\inst{1,}\inst{2,}\thanks{Visiting PhD Student at
  STScI under the \textit{``2008 graduate research assistantship''}
  program.}
\and
Bedin, L.~R.\inst{2}
\and
Piotto, G.\inst{1}
\and
Salaris, M.\inst{3}
\and
Anderson, J.\inst{2}
\and
Brocato, E.\inst{4}
\and
Ragazzoni, R.\inst{5}
\and
Ortolani, S.\inst{1}
\and
Bonanos, A.~Z. \inst{6}
\and
Platais, I.\inst{7}
\and
Gilliland, R.\inst{2}
\and
Raimondo, G.\inst{4}
\and
Bragaglia, A.\inst{8}
\and
Tosi, M.\inst{8}
\and
Gallozzi, S.\inst{9}
\and
Testa, V.\inst{9}
\and
Kochanek, C.~S.\inst{10,11}
\and
Giallongo, E.\inst{9}
\and
Baruffolo, A.\inst{5}
\and
Farinato, J.\inst{5}
\and
Diolaiti, E.\inst{8}
\and
Speziali, R.\inst{9}
\and
Carraro, G.\inst{1,12}
\and
Yadav, R.~K.~S.\inst{13}
}

\offprints{bellini@stsci.edu}

\institute{Dipartimento di Astronomia, Universit\`a di
Padova, vicolo dell'Osservatorio 3, I-35122 Padova, Italy
\and
Space Telescope Science Institute, 3700 San Martin 
Drive, Baltimore, MD 21218, USA
\and
Astrophysics Research Institute, Liverpool
John Moores University, 12 Quays House, Birkenhead, CH41 1LD, UK
\and
INAF - Osservatorio Astronomico di Teramo, via M. Maggini snc,
64100  Teramo, Italy
\and
INAF - Osservatorio Astronomico di Padova, vicolo
dell'Osservatorio 5, 35122 Padova, Italy
\and
Institute of Astronomy and Astrophysics, NOA, I. Metaxa and
Vas. Pavlou Street, Palaia Penteli GR-15236, Greece
\and
Department of Physics and Astronomy, The
Johns Hopkins University, Baltimore, MD 21218, USA
\and
INAF - Osservatorio Astronomico di Bologna, via Ranzani 1, 40127
Bologna, Italy
\and 
INAF - Osservatorio Astronomico di Roma, via Frascati 33, 00040
Monteporzio, Italy
\and
Department of Astronomy, The Ohio State University, 140 West 18th
Avenue, Columbus, OH 43210, USA
\and
Center for Cosmology and Astroparticle Physics, The Ohio State
University, 140 West 18th Avenue, Columbus, OH 43210, USA
\and
ESO, Alonso de Cordova 3107, Vitacura, Santiago de Chile,
Chile
\and
Aryabhatta Research Institute of
Observational Sciences, Manora Peak, Nainital 263129, India
}

\date{Received 23 November 2009 / Accepted 20 january 2010}

\abstract{In this paper, we present for the first time a
  proper-motion-selected white dwarf (WD) sample of the old Galactic
  open cluster M~67, down to the bottom of the WD cooling sequence
  (CS).  The color-magnitude diagram is based on data collected with
  the LBC-Blue camera at the prime focus of LBT.  As first epoch data,
  we used CFHT-archive images collected 10 years before LBC data. We
  measured proper motions of all the identified sources. Proper
  motions are then used to separate foreground and background objects
  from the cluster stars, including WDs.  Finally, the field-object
  cleaned WD CS in the $V$ vs. $B-I$ color-magnitude diagram is
  compared with the models. We confirm that the age derived from the
  location of the bottom of the WD CS is consistent with the turn off
  age.}


\maketitle

\section{Introduction} 

The white dwarf (WD) cooling sequence (CS) lies in one of the most
unexplored parts of the color-magnitude diagram (CMD) of star
clusters.  In a recent deep photometric investigation of the metal
rich open cluster NGC 6791, Bedin et al. (2005;\ 2008a,b) discovered
an unexpectedly bright peak in its WD luminosity function (LF). This
result raises questions about our understanding of the physical
processes that rule the formation of WDs and their cooling phases.  It
is clear that, in order to improve our current understanding of WDs,
in particular at high metallicities, we need to fill the
age-metallicity parameter space of stellar clusters with new data.

Unfortunately, most of the metal-rich old open clusters for which the
end of the WD CS is potentially reachable are relatively sparse.
Because of this, the limited field of view of Hubble Space Telescope
(\textit{HST}) cameras allows us to cover only a small fraction of a
single cluster, which implies that a very limited number of WDs can be
observed.

The dispersion of the cluster stars over a large field has the
additional inconvenience of a strong contamination of all of the
evolutionary sequences in the CMD by foreground/background objects.
The two wide-field imagers (WFIs) at the two prime foci of LBT provide
us with a unique opportunity to overcome the field size problem.
Moreover, the availability of multi-epoch imaging allows us to measure
proper motions, thus alleviating the issue of field contamination.

Note that, although (in principle) it would be possible to entirely
map an open cluster as large as M~67 with \textit{HST}, during its 19
years of activity these kind of projects have never been approved.
For reference, in the case of M~67 it would be necessary a minimum of
$\sim$40 orbits ---per epoch--- to map the inner $\sim$20$\times$20
arcmin$^2$, taking into account for:\ dump-buffer overheads,
intra-orbit pointing limitations, and the necessity to have multiple
exposures with large dithers.  For practical reasons, for nearby
clusters, ground-based telescopes with wide field of view (FoV) are
more appropriate to achieve this purpose, and at much cheaper costs.

In this paper, we present a pioneering work on this subject. For the
first time, we used wide-field astrometry and deep photometry to
obtain a pure sample of WDs in the open cluster M~67 [($\alpha,
\delta)$$_{\rm{J2000.0}}$=(8$^{\rm h}$51$^{\rm
m}$$23\fs3$,$+$11$^\circ$49$^\prime$02${\arcsec}$), Yadav et
al. (2008)].  First epoch photometry from a WFIs is available only
from a 4m class telescope, and our LBC@LBT images were not acquired in
optimal conditions. Yet we have been able to reach the end of the WD
CS, remove field objects by using proper motions, and demonstrate the
potentiality of a wide field imager (particularly if mounted on a 8m
class telescope) for the WD study.

A WD study of M~67 was already published by Richer et al.\
(\cite{richer98}, hereafter R98) using the same data set that we use
here as first epoch images. R98 could not directly see the end of the
WD CS, because of the strong contamination by background field
galaxies, but they could infer its location by a statistical analysis
of star counts around the region in the CMD where the WDs are expected
to be. The great news presented in this paper is that we can now
remove most of the field objects and present a clean WD CS down to its
bottom.

\section{Observations}

We used as first-epoch data a set of images collected on January
10--13, 1997 at the CFHT 3.6m telescope. This data set (a sub-set of
those used also by R98) consists of 15$\times$1200 s $V$-band, and
11$\times$1200 s $I$-band images, obtained with the UH8K camera (8
CCDs, 2K$\times$4K pixels each, in a 4$\times$2 array).  Two chips (\#
2 and 4) of this camera are seriously affected by charge transfer
inefficiency.  As a consequence, the FoV suitable for high precision
measurements is reduced to only six chips (for a total sky coverage of
$\sim$400 arcmin$^2$).  [We will see that none of our WDs fall within
those detectors.]  The median seeing is $1\farcs0$.

\begin{table}[t!]
\small{
\begin{center}
\caption{Log of M~67 data used in this work.}
\label{tab1}
\begin{tabular}{cccc}
\hline\hline &&&\\
$\!\!$\textbf{Filter}&$\!\!$$\!\!$$\!\!$\textbf{\#Images$\times$Exp.\
time}$\!\!$$\!\!$& \textbf{Airmass}&$\!\!$\textbf{Seeing}$\!\!$\\ &
$(s)$ &$(\sec z)$ & (arcsec)\\
&&&\\
\hline
&&&\\
\multicolumn{4}{c}{\textbf{UH8k@CFHT (First Epoch:\ 10--13 Jan 1997)}}\\
&&&\\
$V_{\rm Johnson}$ & $15\times1200$             & 1.02--1.50&0.74--1.37\\
$I_{\rm Johnson}$ & $11\times1200$             & 1.02--1.70& 0.75--1.26\\
&&&\\
&&&\\
\multicolumn{4}{c}{\textbf{LBC@LBT (Second epoch:\ 18 Feb--16 Mar 2007)}}\\
&&&\\
$B_{\rm Bessel}$ & $56\times 180$               & 1.07--1.14 & 0.79--1.88\\
$V_{\rm Bessel}$ & $1\times 15$, $25\times100$& 1.07--1.10 & 0.68--1.26 \\
                & $17\times 110$, $1\times330$& 1.07--1.12 & 0.62--1.31 \\
&&&\\
\hline
\end{tabular}
\end{center}}
\end{table}

The second epoch data, collected between February 16 and March 18,
2007, consists of 56$\times$180 s $B$-band, and 1$\times$15 s,
25$\times$100 s, 17$\times$110 s, 1$\times$330 s $V$-band images,
obtained with the LBC-blue camera (4 CCDs, 2K$\times$4.5K pixels each,
3 aligned longside, one rotated 90$^\circ$ on top of them, FoV of
$\sim$24$^\prime$$\times$25$^\prime$, see Giallongo et al.\
\cite{giallongo08}).  This data set is not optimal:\ the original
project consists in a set of 25$\times$180s $B$ and 25$\times$110s $V$
images for science purposes, plus 25$\times$100s $V$ images to solve
for the geometric distortion (hereafter GD).  All 100 s $V$ images
have anomalously high background ($\sim$20$\,$000 digital numbers
(DNs) instead of an expected $\sim$3000).  The median seeing is
$1\farcs0$ for the $V$ filter, and $1\farcs3$ for the $B$ one.  See
Table 1 for the complete log of observations.

\section{Measurements and Selections}

We successfully exported our data reduction software developed for
\textit{HST} images (Anderson et al.\ \cite{anderson08}), adapting it
to the case of ground-based WFIs.  Below, we briefly describe our
3-step procedure, used on both LBC and UH8K data sets.  Further
details on data reduction and calibration procedures are presented in
two companion papers (Bellini \& Bedin submitted, Bellini et
al.\ in preparation).

\begin{figure}[t!]
\centering 
\includegraphics[width=\columnwidth]{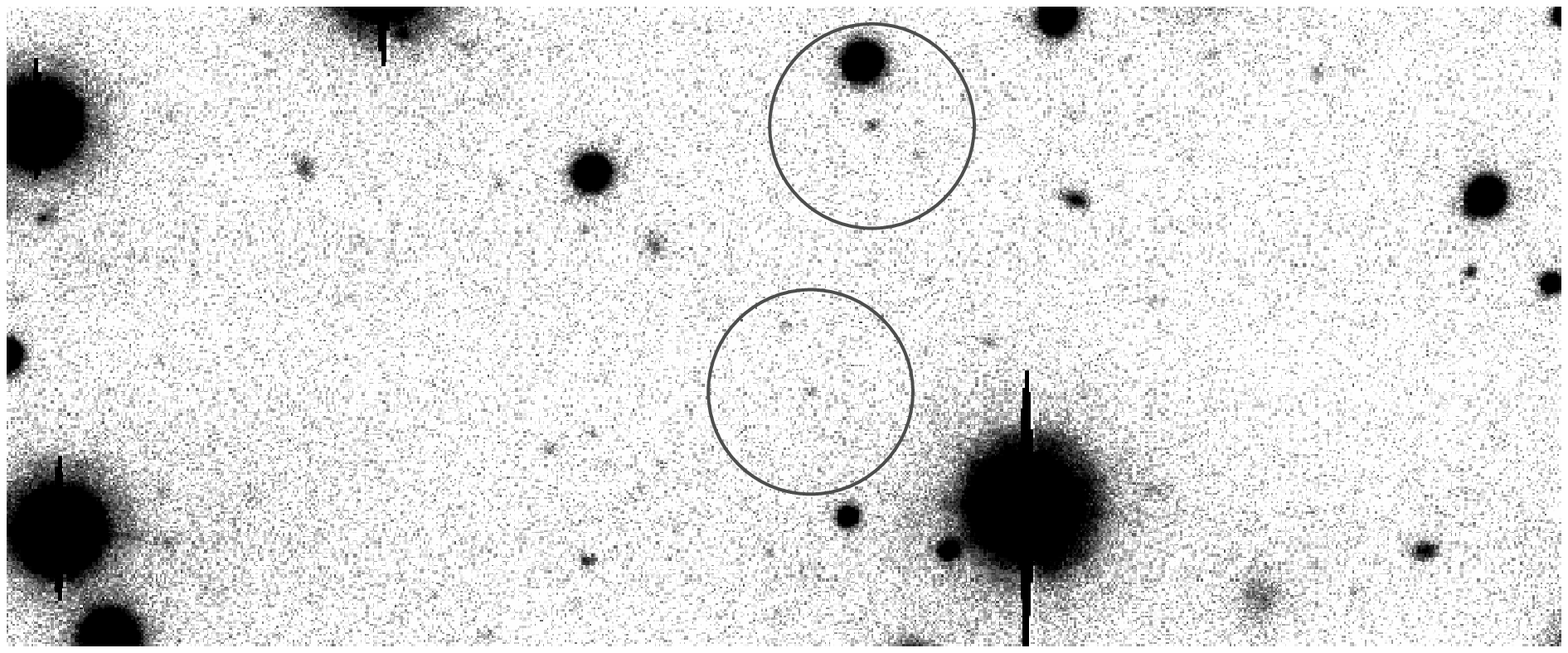}\\
\includegraphics[width=\columnwidth]{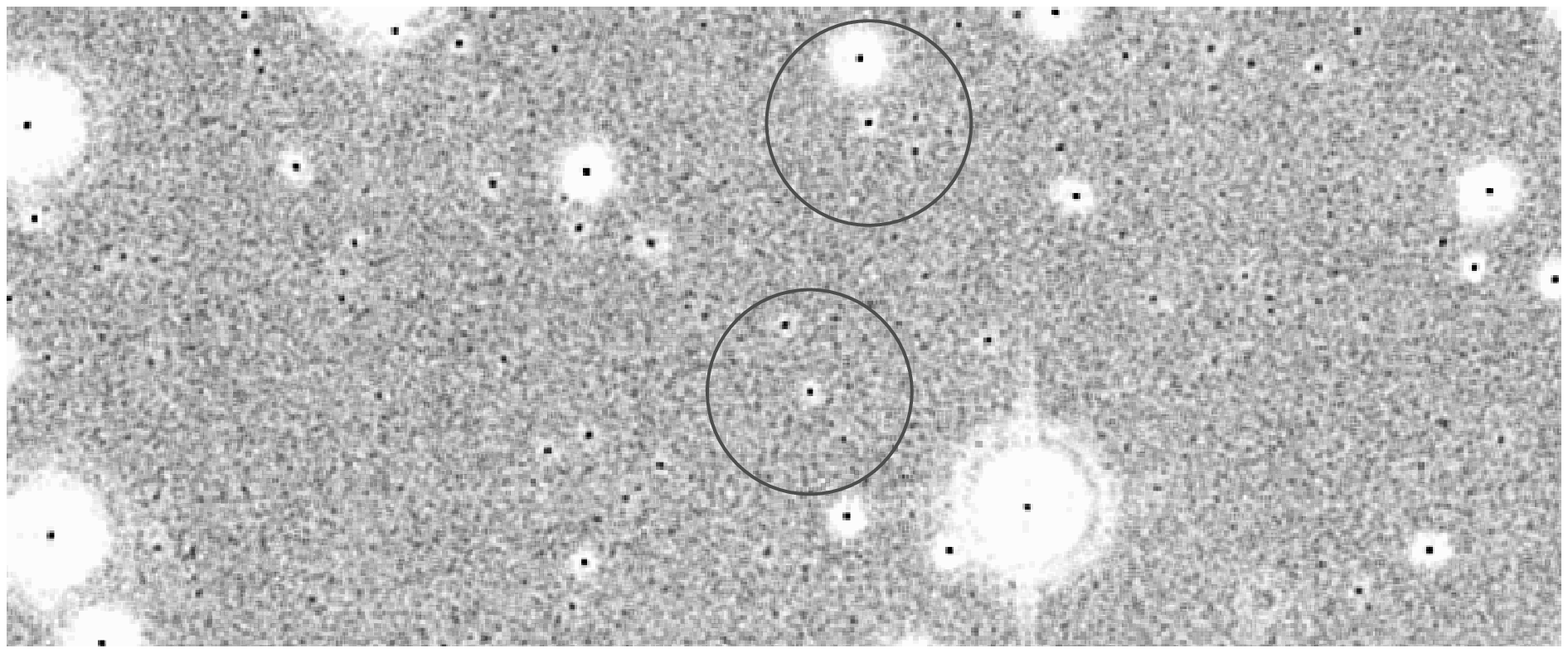}\\
\caption{\textit{(Top:)} A region of a LBT $V$ image (110 s exposure,
seeing $\sim$1$\arcsec$, North is up, East is to the left) in which
two WDs are present, highlighted by open circles (the southern-most
one is also the faintest WD in our sample). \textit{(Bottom):} the
same region as seen on our 3$\times$3 overbinned peak map. The FoV is
$\sim$170$\arcsec$$\times$70$\arcsec$.  See text for details.}
\label{f:pkm}
\end{figure}

\begin{figure*}[t!]
\centering
\includegraphics[width=9cm]{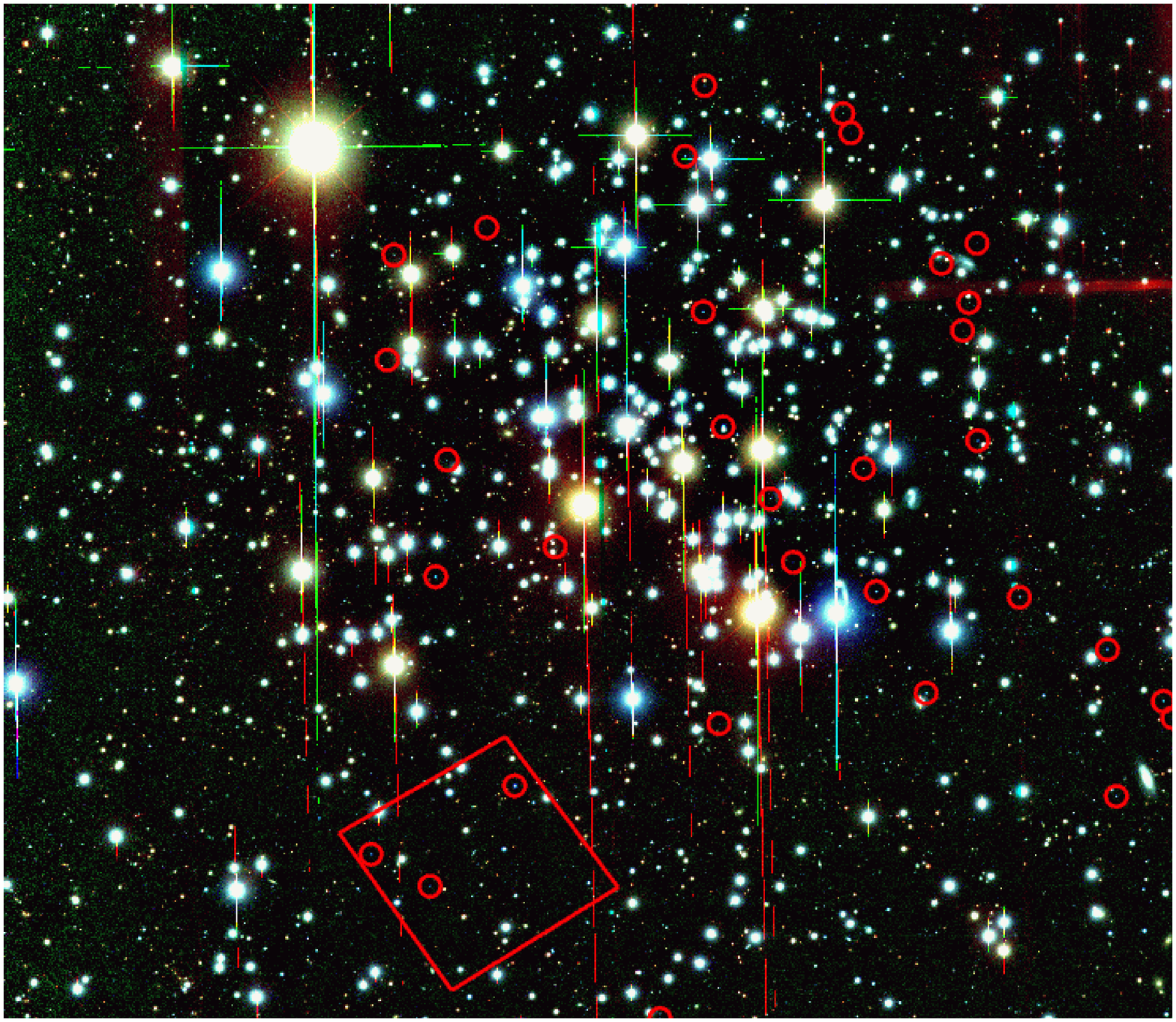}
\includegraphics[width=9cm]{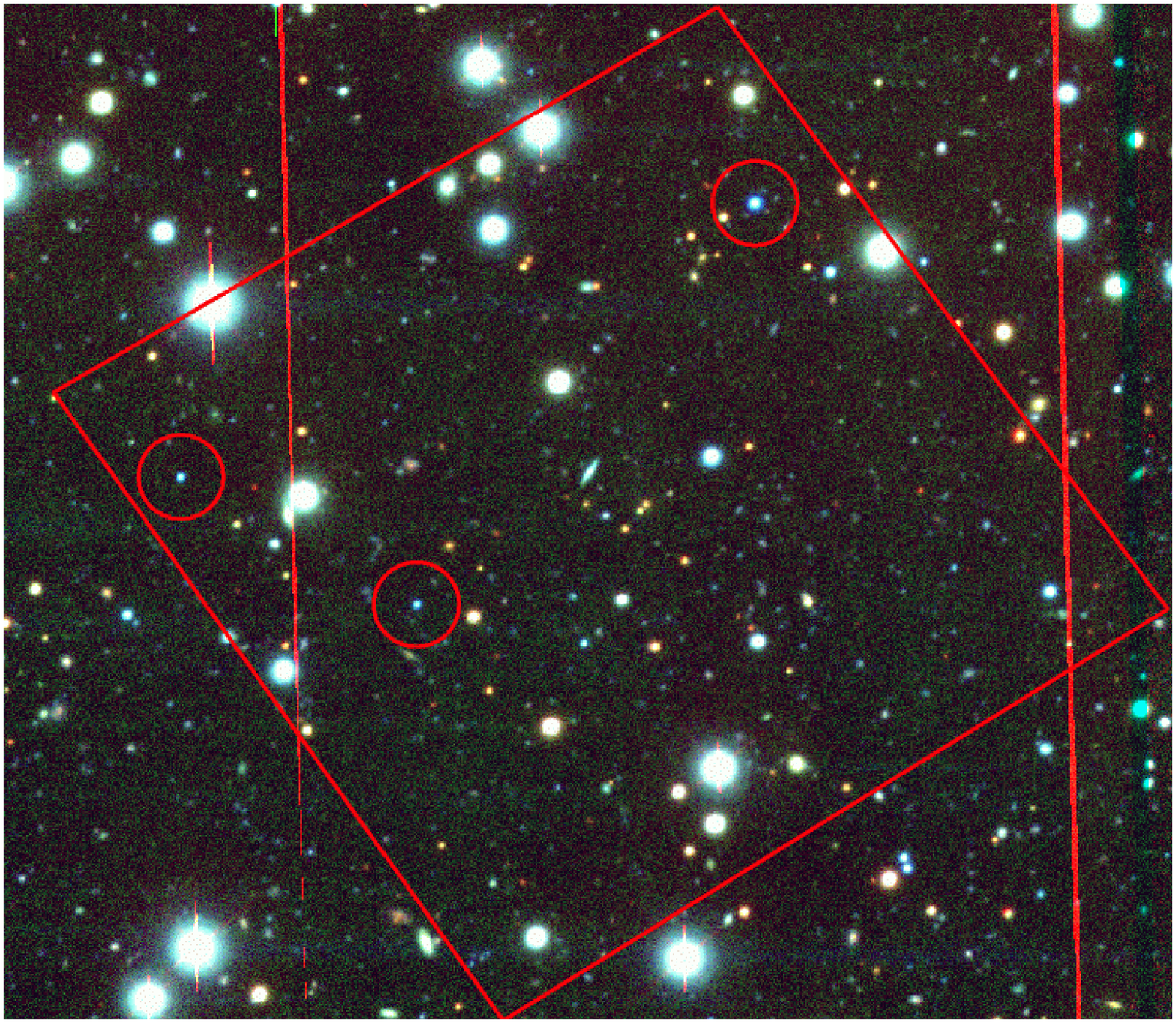}\\
\includegraphics[width=9cm]{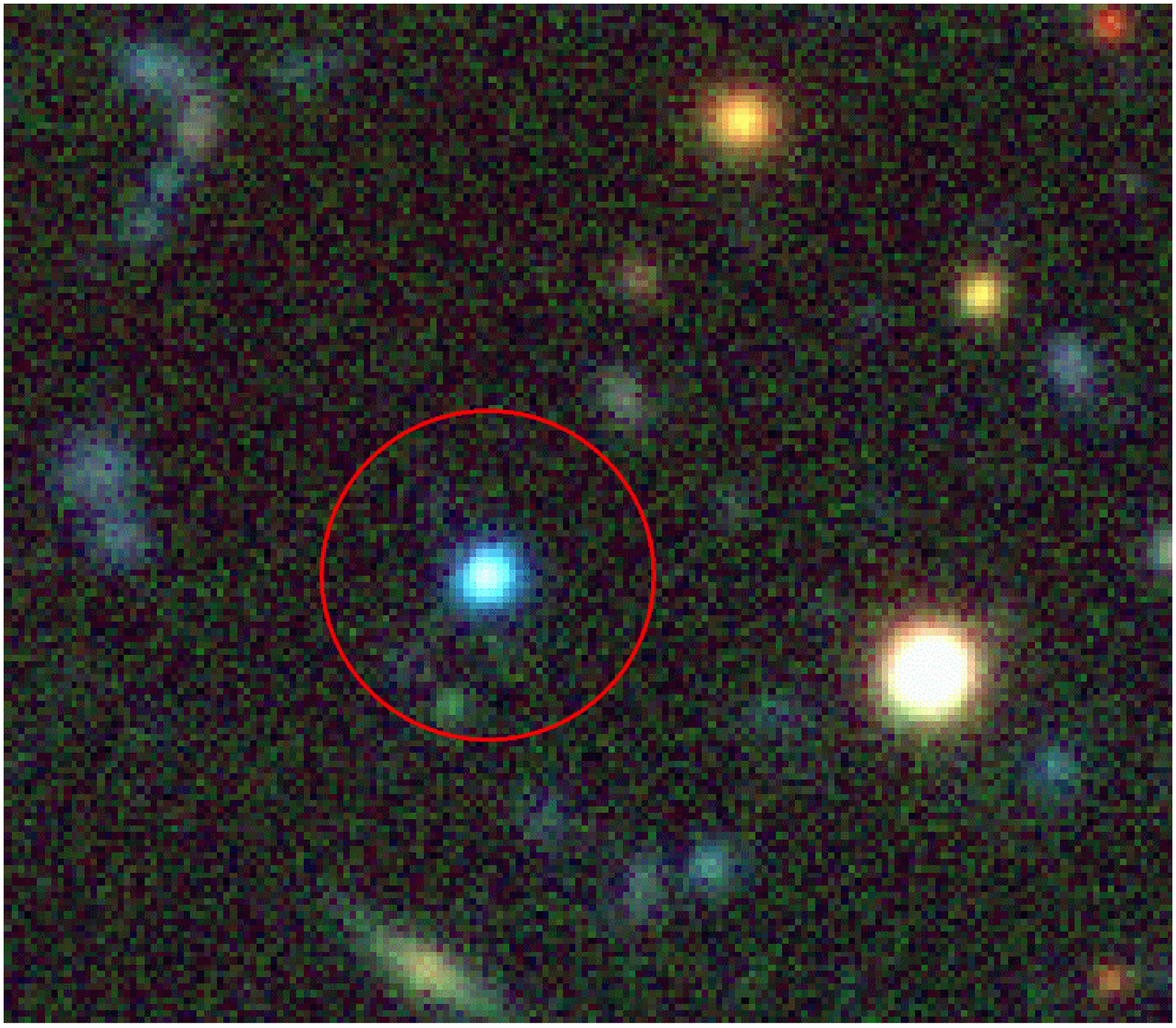}  
\includegraphics[width=9cm]{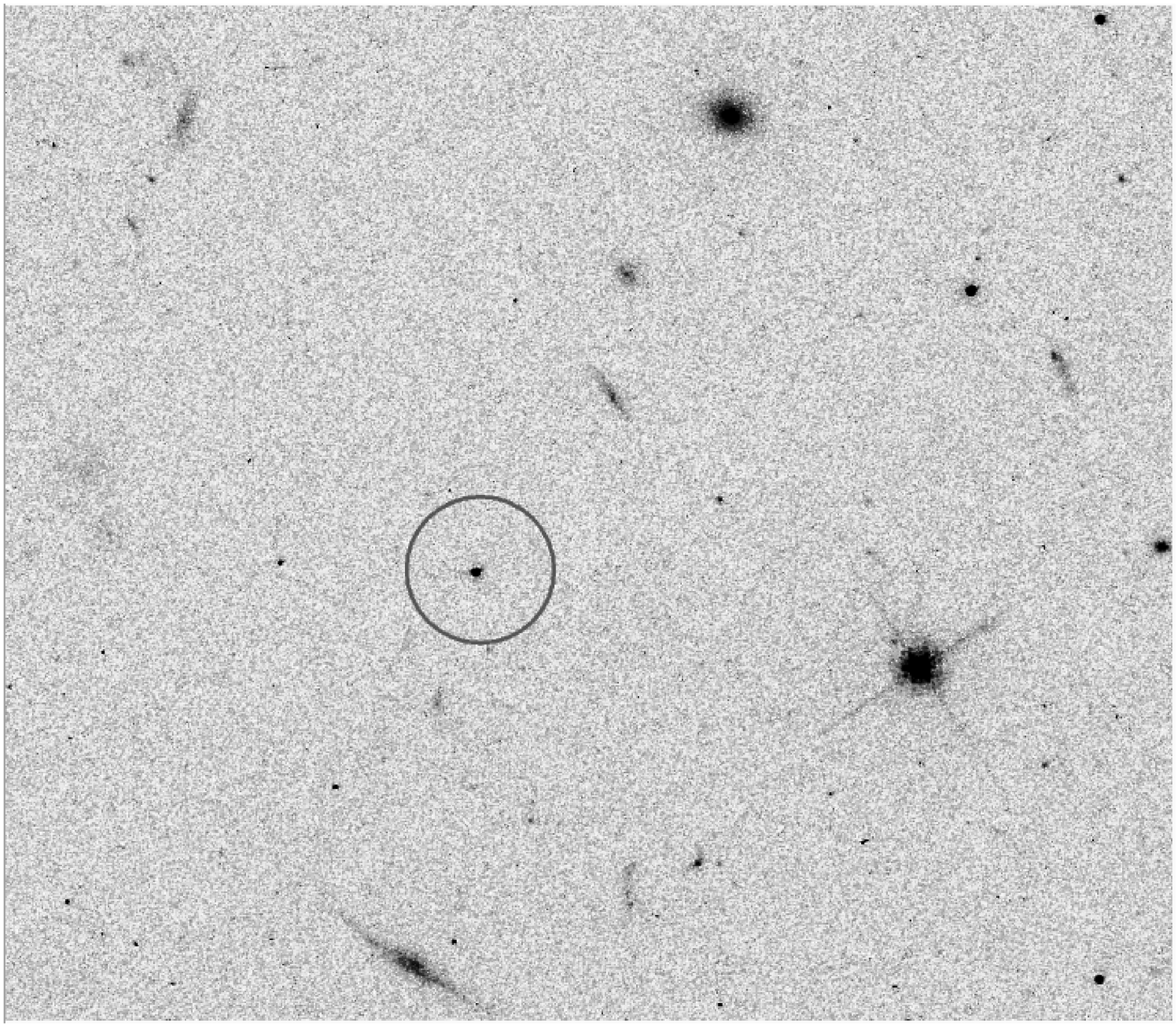} 
\caption{{\it (Top left):} The central $\sim$20$\times$17 arcmin$^2$
region of our stacked trichromatic image from $BV$-LBT and $I$-CFHT
frames.  North is up, East is to the left.  Red circles highlight the
proper-motion selected WDs within the region. The red trapezoid
indicates an {\it HST} ACS/WFC field from the archive (combining
1$\times$400 s and 1$\times$568 s images in filter F775W; GO-9984).
{{\it (Top right):}} An enlargement of the area in common with the
ACS/WFC field.  {\it (Bottom left):} Zoom-in around the southern-most
WD within the ACS/WFC field. The area is $\sim$40$\arcsec$ wide.  The
circle has a radius of 25 pixels.  {\it (Bottom right):} The same
portion of the field, but as seen in the ACS/WFC stacked image.}
\label{f:FoV}
\end{figure*}

\subsection{First step}

We used the software described in Anderson et al.\ (\cite{anderson06})
to obtain spatially varying empirical PSFs (in an array of 3$\times$7
each chip for LBC, and in an array of 3$\times$5 for each UH8K chip),
with which we measured star positions and fluxes in each single chip
of each individual exposure.  We then corrected LBC raw positions for
GD according to Bellini \& Bedin (submitted). In a nutshell, we
modeled the GD with a third-order polynomial, for each LBC chip
independently. For each $i$-star in each $k$-chip, the distortion
corrected position $(x_{i,k}^{\rm corr},y_{i,k}^{\rm corr})$ is
obtained as the observed position $(x_{i,k},y_{i,k})$ plus the
distortion correction $(\delta x_{i,k},\delta y_{i,k})$:
$$
\left\{
\begin{array}{rcl}
x_{i,k}^{\rm corr}&\!\!=\!\!&x_{i,k}+\delta
x_{i,k}(\tilde{x}_{i,k},\tilde{y}_{i,k})\\
y_{i,k}^{\rm corr}&\!\!=\!\!&y_{i,k}+\delta
y_{i,k}(\tilde{x}_{i,k},\tilde{y}_{i,k}),\\
\end{array}
\right.
$$
where $\tilde{x}_{i,k}$ and $\tilde{y}_{i,k}$ are normalized positions
with respect to the center $(x_\circ,y_\circ)_k$ of the $k$-chip
[assumed to be the pixel (1025,2305) for chips \# 1 to 3, and
(2305,1025) for chip \# 4], and $(\delta x_{i,k},\delta y_{i,k})$ are
given as:
$$
\left\{
\begin{array}{rcl}
\displaystyle\!\!  \delta
x&\!\!=\!\!&a_1\tilde{x}\!+\!a_2\tilde{y}\!+\!  a_3\tilde{x}^2 \!+\!
a_4 \tilde{x}\tilde{y}\!+\!  a_5\tilde{y}^2 \!+\!
a_6\tilde{x}^3\!+\!a_7\tilde{x}^2\tilde{y} \!+\!
a_8\tilde{x}\tilde{y}^2 \!+\!  a_9\tilde{y}^3\\ 
\displaystyle \!\!
\delta y&\!\!=\!\!&b_1\tilde{x}\!+\!b_2\tilde{y}\!+\!  b_3\tilde{x}^2
\!+\!  b_4 \tilde{x}\tilde{y}\!+\!  b_5\tilde{y}^2 \!+\!
b_6\tilde{x}^3\!+\!b_7\tilde{x}^2\tilde{y} \!+\!
b_8\tilde{x}\tilde{y}^2 \!+\!  b_9\tilde{y}^3\\
\end{array}
\right.
$$
(where we omitted here the subscript ``${i,k}$'' for
simplicity).

The GD solution is therefore fully characterized by 18 coefficients:
$a_1,\dots, a_9$, $b_1,\dots, b_9$.  We constrained the solution so
that, at the center of the chip, it will have its $x_k$-scale equal to
the one at the location $(x_\circ,y_\circ)_k$, and the corrected axis
$y^{\rm corr}_k$ has to be aligned with its $y_k$-axis at the location
$(x_\circ,y_\circ)_k$. This is obtained by imposing $a_{1,k}=0$ and
$a_{2,k}=0$, so we are left with only 16 coefficients.  Our average
correction enables a relative astrometric accuracy of $\sim$10 mas per
coordinate (see Bellini \& Bedin for a detailed description of the GD
solution derivation, and tables with the polynomial coefficients).

We then used these GD-corrected positions to register all of the LBC
single chips into a common distortion-free frame (the master frame),
using linear transformations.  The transformations were computed using
the best sources in the field (bright, isolated, and with a stellar
profile).  With this GD-corrected master frame we derived the GD
correction also for the UH8K camera, by means of the same technique
(used also in Bellini \& Bedin \cite{bb09}), and we corrected and
registered all first- and second-epoch positions to the LBC master
frame.

\subsection{Second step}

In the second step, we used every single local maximum detected within
each single chip of each LBC $B$-image to build a peak-map image,
i.e.\ a map of how often a local maximum occurred at a particular
place in the field. A local maximum (peak) is any pixel of an image
whose flux is strictly higher than any of its eight surrounding
neighbors.  We used $B$ images for the construction of the peak map
because they allowed us to detect the faintest sources. This peak map
consists of an image with the same average pixel scale of the master
frame, where we added 1 to a pixel count each time a local maximum,
measured in a given image, fell within that pixel (once transformed
with the aforementioned linear transformations).

A 3$\times$3 box-car filter is applied to the peak map (as done in
Anderson et al.\ \cite{anderson08}), in order to optimally highlight
the signal from the faintest objects.  This overbinning was necessary
because very faint sources do not necessarily fall within the same
pixels (due to noise fluctuations, the brightest pixel of a faint
source might not be at the location predicted by the PSF, but instead
in one of its neighbor pixels), and we want to consider all of the
peaks that each source generates in order to maximize the signal.

In the top panel of Fig.\ \ref{f:pkm} we show a region of the FoV
($\sim$750$\times$300 pixel$^2$, $\sim$170$\times$70 arcsec$^2$) as
imaged by a LBT $V$ image of 110 s (seeing $\sim$1$\arcsec$).  North
is up, East to the left. The two sources marked with open circles are
WDs, and the southern-most of the two, magnitude $V$=24.1, is the
faintest M~67 WD detected in our sample (see Section
\ref{ss:3.4}). These two stars are barely visible in the single
exposure, but they are easily detectable in the 3$\times$3 overbinned
peak map (bottom panel of Fig.\ \ref{f:pkm}). The faintest WD is able
to generate a peak in 49/56 $B$ exposures.

The peak map is then used to generate our faint-source list (the
target list).  We adopted a two-parameter algorithm:\ In order to be
included in our initial list, a source has to (1) generate a peak
which value is at least 60\% of the number of exposures mapping its
location, and (2) be at least 5 pixels away from any more significant
source.  These criteria correspond, on average, including 3.2$\sigma$
detections above the local surrounding (where $\sigma$\ is the rms of
the peak-map background; see Anderson et al.\ 2008 for more details).
Lowering the threshold below this limit would dramatically increase
the noise contribution to our sample.  Our target list, does not
contain sources located in patches of sky that were covered by less
than 10 $B$ images.

It is useful to construct stacked images so that we can examine the
images of stars and galaxies that are hard to see clearly in
individual exposures. The stacks (one each filter) are invaluable in
helping us discriminate between real objects the software should
classify as stars or galaxies, and objects that should be rejected as
PSF bumps or artifacts.  We used the positional transformations
between each chip and the master frame to determine where each pixel
in each exposure mapped onto the stack frame.  The value of these
pixels (sky subtracted) was properly scaled to match the photometric
zero point of the master frame.  For a given pixel of the stack image,
we can dispose of several such mapping pixels.  We assigned to a stack
pixel the 3$\sigma$-clipped median of its mapping pixels, where
$\sigma$ is rms of the residuals around the median.  Unlike Anderson
et al.\ (\cite{anderson08}), we did not iterate this procedure, since
both LBC and UH8K cameras are not undersampled.

Top-left panel of Figure\ \ref{f:FoV} shows the trichromatic stacked
image (RGB color-coded with $I$, $V$, $B$, respectively) of the
central region of M~67.  Red circles mark M~67 WD members (see Section
\ref{ss:3.4}). The trapezoid delineates the patch of sky in common
with an archive Advanced Camera for Survey (ACS) Wide Field Channel
(WFC) \textit{HST} field, from GO-9984. North is up, East to the
left. The top-right panel shows a zoom-in around the region in common
with ACS/WFC; on the bottom-left a closer view of the southernmost WD
in previous panel;\ finally, the same region as it appears in an
ACS/WFC stack (F775W filter, total exposure time $\sim$1070 s). It is
clear that our stacked image is able to reveal as many objects as the
ACS/WFC one.

More importantly, in our stacked images even the faintest WDs stand
out well above the surrounding background noise. In the top-left panel
of Fig.\ \ref{f:cmd_pm} we show a 40$\arcsec$$\times$40$\arcsec$ region of
the $B$-stacked image, centered around the faintest M~67 WD measured
in this work (highlighted with an open circle).  This star is clearly
visible in our stacked image. Its brightest pixel has $\sim$75 DNs
(sky subtracted) above a local sky noise of $\sim$4 DNs, therefore its
detection is unambiguous. This star (that has a magnitude of
$V$=24.1), is surrounded by many other fainter sources, the majority
of which are background galaxies (note their more asymmetric and
blurred shape, if compared to the WD).

\subsection{Third step}
\label{ss:3.3}

We collected all the information for a given source, in each filter
individually, following the prescriptions given in Anderson et al.\
(\cite{anderson08}), and described here below. We used the positional
transformations and the distortion corrections to calculate the
position where each source of our target list falls in each individual
exposure, and extracted a 11$\times$11 array of pixels (raster) around
the predicted position.  Since the chips have different zero points,
and the image quality is different from one image to the other, we
corrected each raster (sky subtracted) to the proper photometric zero
point of the master frame.

Our local PSF model tells us the fraction of star light that is
expected to fall in a pixel centered at an offset $(\Delta x,\Delta
y)$ from the star's center, for any given image. Therefore, the flux
$P$ in each pixel $(i,j)$ in each raster $n$ is described by:
\begin{equation}
~~~~~~~~~~~~~~~~
~~~~~~~~~~~~~~~~P_{i,j,n}-s_{\ast,n}=f_{\ast}\cdot\psi_{i,j,n},
\label{eq:1}
\end{equation}
where $f_{\ast}$ is the star's flux, $s_{\ast,n}$ is the local
background value, and $\psi_{i,j,n}$ is the fraction of light that
should fall in that pixel, according to the local PSF model. This is
the equation of a straight line with a slope of $f_{\ast}$ and a null
intercept (note that our rasters are already sky subtracted).  We fit
the flux $f_{\ast}$ for each star by a least-squares fit to all the
pixels within $r$$<$5 pixel from the star's center, taking into
account the expected noise in each pixel.

Unlike Anderson et al.\ (\cite{anderson08}), we did not calculate a
further local sky value prior to solving for $f_{\ast}$ in
Equation~\ref{eq:1}, because star light is spread all over the raster.
We performed an additional procedure instead.  After we had a first
$f_{\ast}$ value, we went back into every single raster and we
subtracted the quantity $f_{\ast}\cdot\psi_{i,j}$ from each pixel
value $P_{i,j}$. We used the 3$\sigma$-clipped average of the 40
pixels between $r$=3.5 and $r$=5 to estimate the background residual
${\rm res}_{\ast}$.  Finally, we recalculated the star's flux
$f_{\ast}$ by solving the new equation
\begin{equation}
~~~~~~~~~~~~~~~
~~~~~~~~~~~~~~~~
f_{\ast}=\frac{P_{i,j,n}-s_{\ast,n}-{\rm res}_{\ast,n}}{\psi_{i,j,n}}.
\label{eq:2}
\end{equation}
We iteratively rejected the points that were more than 3$\sigma$
discordant with the best-fitting model.

\begin{figure}[t!]
\centering
\includegraphics[width=9cm]{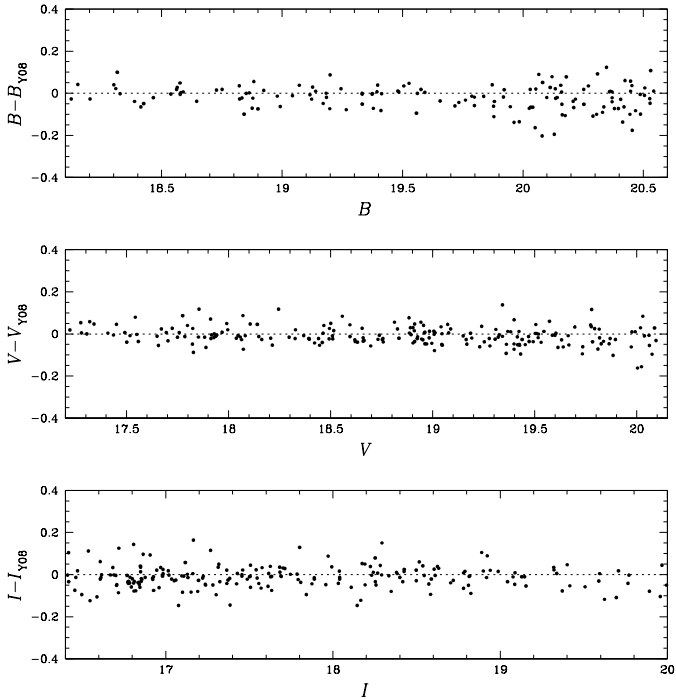}
\caption{Magnitude differences between our photometry and the one
presented in Yadav et al.\ (\cite{yadav08}) catalog (indicated here
with the subscript Y08) for the sources in common. From top to bottom,
$B-B_{\rm Y08}$ vs. $B$, $V-V_{\rm Y08}$ vs. $V$, $I-I_{\rm Y08}$
vs. $I$, respectively.}
\label{f:cal}
\end{figure}

We fit always more than 790 individual-pixel values (cfr.\ Sect.\ 4 in
Anderson et al.\ \cite{anderson08}), and up to $\sim$4400.  This has
been done independently for each filter at each epoch, using the same
star list.  The uncertainty of the slope is the formal error of the
least-squares fit, and provides our internal estimate of the
photometric error.  We will explain in Sec.~\ref{thWD} how to obtain a
more reliable external estimate of the true errors.  The flux is then
converted into instrumental magnitudes, and calibrated to the $BV$
Johnson $I$ Kron-Cousin standard systems, using as secondary standards
the objects from the Yadav et al.\ (\cite{yadav08}) catalog.  In Fig.\
\ref{f:cal} we show, for common sources, the differences in magnitude
between our $B$, $V$, $I$ unsaturated photometry and the one published
in Yadav et al.\ (\cite{yadav08}).

\begin{figure*}[t!]
\begin{tabular}{ll}
\centering
\includegraphics[width=4.500cm]{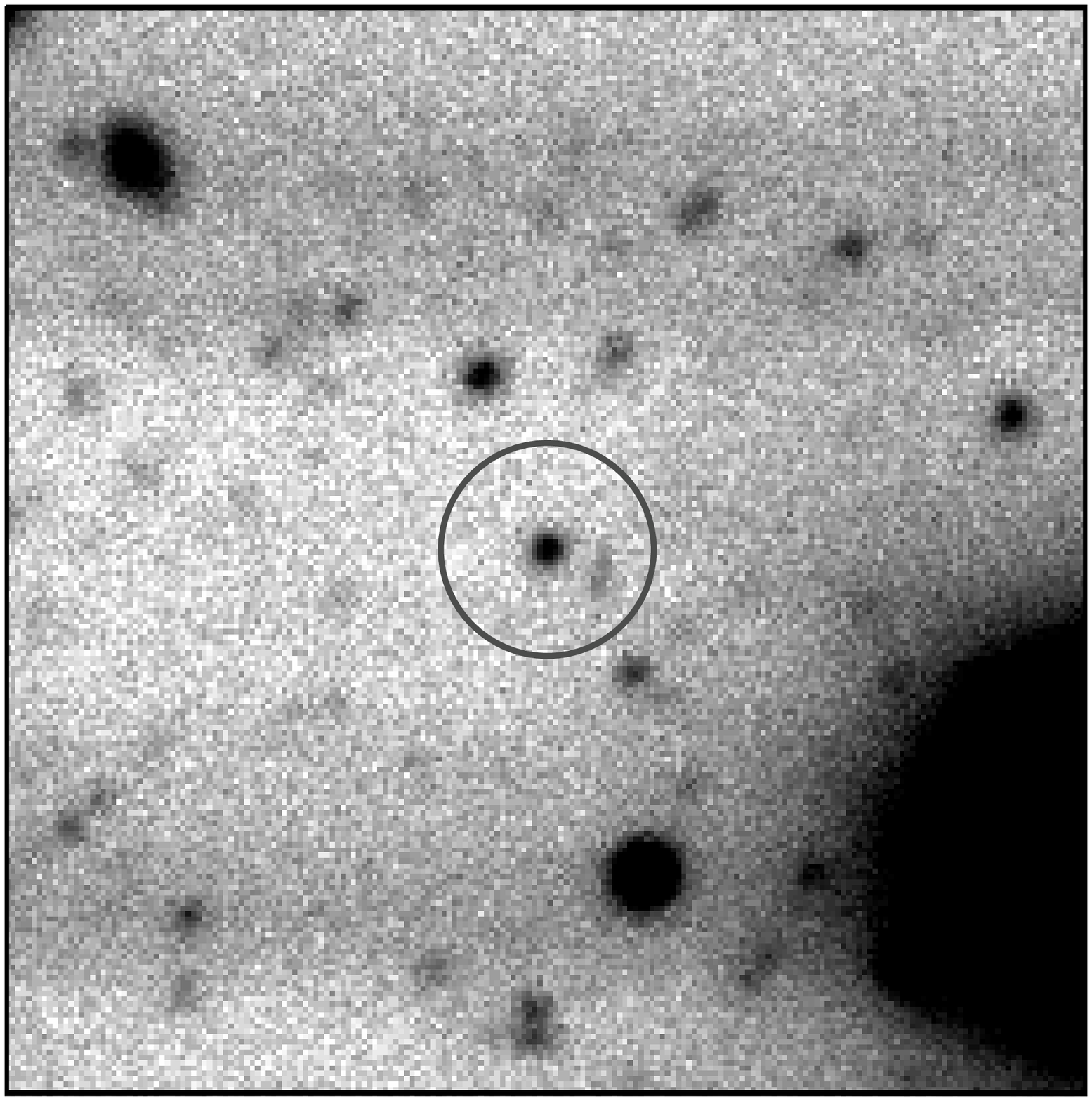}
\includegraphics[width=17.75cm]{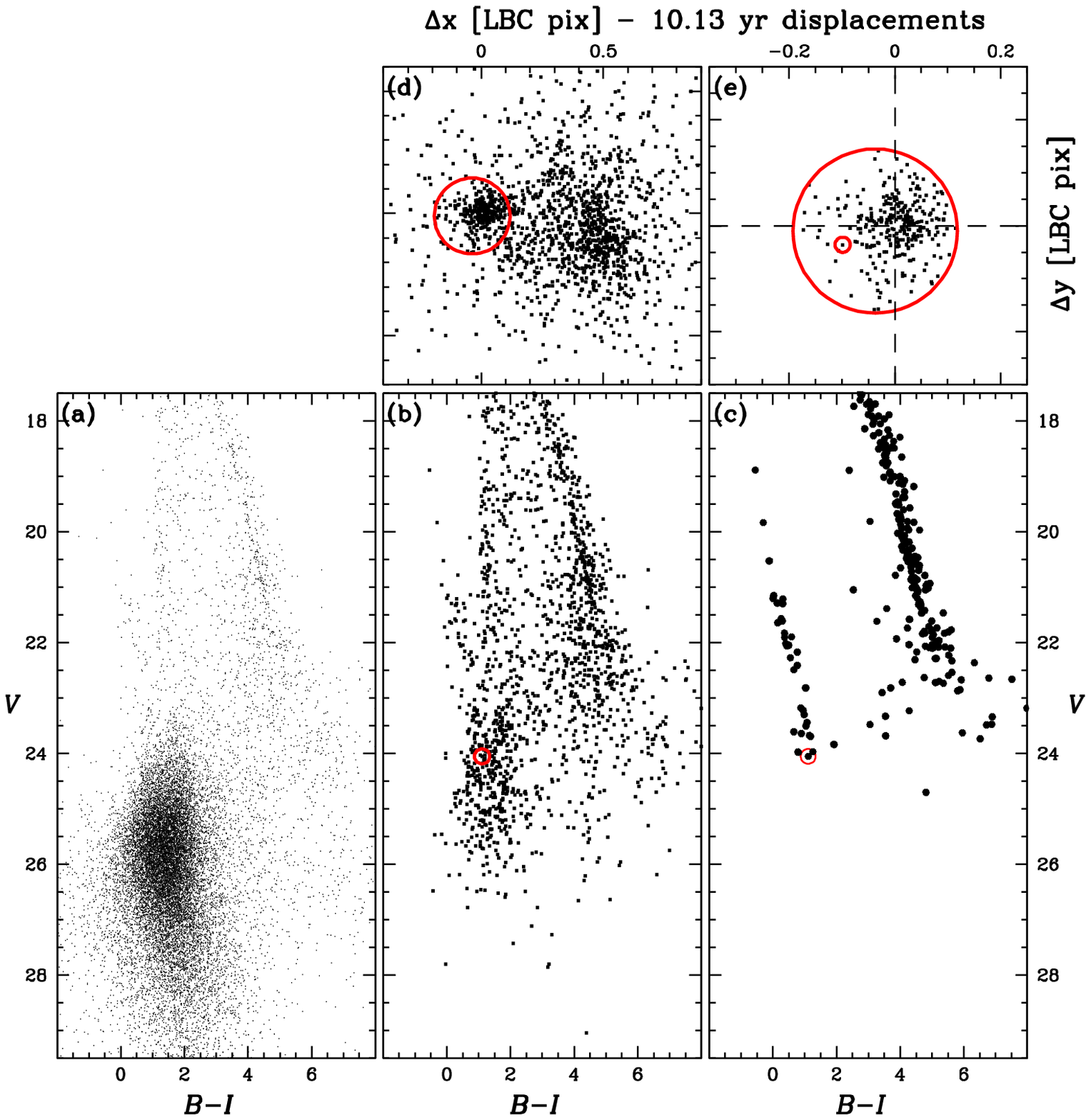}
\end{tabular}
\caption{Panel (a) shows the $V$ vs. $B-I$ CMD for all the objects in
the target list.  With our photometric-reduction technique we are able
to measure sources as faint as $V$$\gtrsim$28. It is clear from the
figure that the bulk of faint blue compact galaxies occupy the same
region as the WDs.  Panel (b) shows the same CMD for objects for
which we have a proper motion measurement.  Our proper motions can
reach a magnitude of $V$$\sim$26.  The relative VPD is shown in panel
(d);\ a red circle includes the objects that we assumed to be cluster
members (see text for details).  Panel (c) shows the CMD for the
cluster members only, as selected in panel (d).  Panel (e) shows the
zoomed VPD for members.  A small red circle in panels (b), (c), (e)
identifies the faintest WD.  In order to show the significance of the
detection of this WD, top-left panel shows a
40$\arcsec$$\times$40$\arcsec$ region of the $B$-stacked image around
this star, that stands clearly above the background.  The brightest
pixel of this star has $\sim$75 DNs (sky subtracted) above a local sky
noise of $\sim$4 DNs, therefore its detection is unambiguous.}
\label{f:cmd_pm}
\end{figure*}

In panel (a) of Fig.\ \ref{f:cmd_pm} we show the $V$ vs. $B-I$ CMD
of all the sources measured this way. Our photometric-reduction
techniques allow us to measure the faintest sources in our data set
($V$$\geq$28). The M~67 main sequence (MS) and WD CS are embedded in a
large number of foreground and background objects which prevent us from
seeing the end of both sequences.  In particular, we will show that the
vast majority of sources forming the dense clump at $V$$>$23.5 and
$B-I$$<$3 in the CMD are faint blue compact galaxies.  This clump is
almost exactly where we expect to find the bottom of the WD CS.  In the
next section, we will use proper motions to remove field sources from
the CMD and isolate M~67 stars.

\subsection{Proper motions}
\label{ss:3.4}

Proper motions were obtained as described in Anderson et al.\
(\cite{anderson06}).  Of the initial target list obtained from the
peak map, we selected only those sources with a measure in both LBC
$V$ and $B$ filters.  This list is then used to calculate
displacements between the two epochs, as follows.

We measured a chip-based flux and GD-corrected position using
PSF-fitting for each object in the list in each chip of each exposure
where it could be found.  Then we organized the images in pairs of
images from each epoch.  For each object, in each pair, we computed
the displacement (in the reference frame) between where the
first-epoch predicts the object to be located, after having
transformed its coordinates on the reference-frame system, and its
actual-observed position in the second epoch image.  Multiple
measurements of displacements for the same object are then used to
compute average displacements and rms.

It is clear that, in order to make these displacement predictions, we
need a set of objects to be used as a reference to compute positional
transformations between the two epochs for each source.  The cluster
members of M~67 are a natural choice, as their internal motion is
within our measurement errors ($\sim$0.2 mas yr$^{-1}$ Girard et al.\
\cite{girard89}), providing an almost rigid reference system with the
common systemic motion of the cluster.

We initially identified cluster members according to their location in
the LBC $V$ vs.\ $B-V$ color-magnitude diagram.  By predominantly
using cluster members, we ensure proper motions to be measured
relative to the bulk motion of the cluster.  We iteratively removed
from the member list those objects with a field-type motion even
though their colors may have placed them near the cluster sequences.
In our proper-motion selections (see panel (d) of
Fig.\ \ref{f:cmd_pm}), we considered as cluster members all sources
whose displacement is within a circle of radius 0.15 LBC pixels in
10.13 yr [we found this to be the best compromise between losing
  (poorly measured) M~67 members and including field objects] in the
  vector-point diagram (VPD), slightly off-center with respect to the
zero of the motion, in order to further reduce the number of field
objects in our member list.

In order to minimize the influence of any uncorrected GD residual,
proper motions for each object were computed using a local sample of
members; specifically the 25 (at least) closest ($r$$<$3$^\prime$),
well-measured cluster stars (see Anderson et al.\ \cite{anderson06}
for more details).  Note that, in order to maximize the cluster-field
separation, we used all the available images, in every filter, in both
epochs. Indeed, our aim here is to provide a pure sample of M~67 WDs,
and we are not interested in removing systematic errors below the mas
level at the expenses of the size of the WD sample.  For a more
careful proper motion analysis we refer the reader to the companion
paper Bellini et al.\ (submitted).

Finally, we corrected our displacements for atmospheric differential
chromatic refractions (DCR) effects, as done in Anderson et al.\
(\cite{anderson06}).  The DCR effect causes a shift in the photocenter
of sources, which is proportional to their wavelength, and a function
of the zenithal distance: blue photons will occupy a position that
differs from that of red photons.  Unfortunately, within each epoch,
the available data sets are not optimized to perform the DCR
correction directly.  We can, however, check if possible differences
in the DCR effects between the two epochs could generate an apparent
proper motion for blue stars relative to red stars.

We selected two samples of cluster member stars, one made by WDs in
the magnitude interval 20$<$$V$$<$23, and the other on the MS, within
the same magnitude interval. For each sample, we derived the median
$B-V$ color and the median proper motion along the $X$ and $Y$ axes,
with dispersion obtained as the 68.27$^{\rm th}$ percentile around the
median.  We used a linear fit to derive the DCR corrections along both
axes, as done in Anderson at al.\ (\cite{anderson06}).  We found that
DCR corrections were always below 0.7 mas yr$^{-1}$ $(B-V)^{-1}$.

In panel (b) of Fig.\ \ref{f:cmd_pm} we show the $V$ vs. $B-I$ CMD of all
the sources for which we have at least two individual displacement
measurements.  The red circle in the figure marks the location of the
faintest M~67 WD measured in this work. It is clear from panel (b)
that we can measure proper motions of sources more than two magnitudes
fainter than this star. Panel (d) in the same figure shows the VPD of
the sources in panel (b). Since we used M~67 star members as reference
to compute displacements between the two epochs, the origin of the
coordinate coincides with the M~67 mean motion. The red circle in
panel (d) show our adopted membership criterion.

The cluster-field separation is $\sim$0.5 pixel in 10.13 yr ($\sim$11
mas$\,$yr$^{-1}$).  This separation is consistent with the one
presented in Yadav et al. (\cite{yadav08}).  A detailed comparison,
and the merged catalogs, will be given in a companion paper and will
deliver to the community the ---so far--- most complete catalog of
M~67 members.

The $V$ vs. $B-I$ CMD of the selected cluster members is shown in
panel (c), and the corresponding VPD is shown (enlarged for a better
reading) in panel (e). Again, we highlighted the position of the
faintest measured M~67 WD with a small circle, in both panels.  It is
clear from panel (c) that the WD CS sharply drops at
$V$$\sim$24.1. This magnitude marks the bottom of the WD CS of M~67.
The faintest WD has a proper motion consistent with the M~67 mean
motion [see panel (e)].  A visual inspection of all the M~67 WDs on
our stacked images confirmed that all of them are real stars.

\subsection{Completeness}
\label{ss:3.5}

A complete investigation of the WD CS would require also the study of
the WD LF. However, a proper study of the LF requires an appropriate
estimate of the completeness of the star counts.

Computing completeness corrections taking into account both photometry
and proper motions is a very complex and delicate matter, which is
beyond the capability of our software, which was developed with the
specific aim of deriving the most precise photometry and proper
motions.  A completeness correction that takes carefully into account
proper motions opens a new series of problems, has never been properly
treated in the literature.  The aim of the present paper is to extract
a clean sample of M~67 WD members to precisely locate them on the CMD,
to study their main properties and to constrain theory.  Our catalog
will provide observers with a set of bona-fine WDs for spectroscopic
follow-up investigations (at least for the bright WDs).  We located
the bottom of the WD CS just where the CS ends, not using statistical
subtractions in the LF, as done, e.g. by R98 on the same cluster.
However, it is important to emphasize here that our photometry and
astrometry extends far below (about two magnitudes at the color of the
faintest WD) the sharp WD CS cut off. Moreover, as shown by R98, at
the $V$$<$25, the completeness, even in the shallower CFHT images, is
$>$95\%. We do not expect that our reduction software and the proper
motion selection procedure can lower this completeness in any
significant way.  We defer the analysis of the completeness to when a
new set of LBC images (project already approved by LBT TAC) will be
available, providing a third epoch for a much more accurate proper
motion measurement.

\section{Comparison with previous studies}

\begin{figure*}[t!]
\centering
\includegraphics[width=18cm]{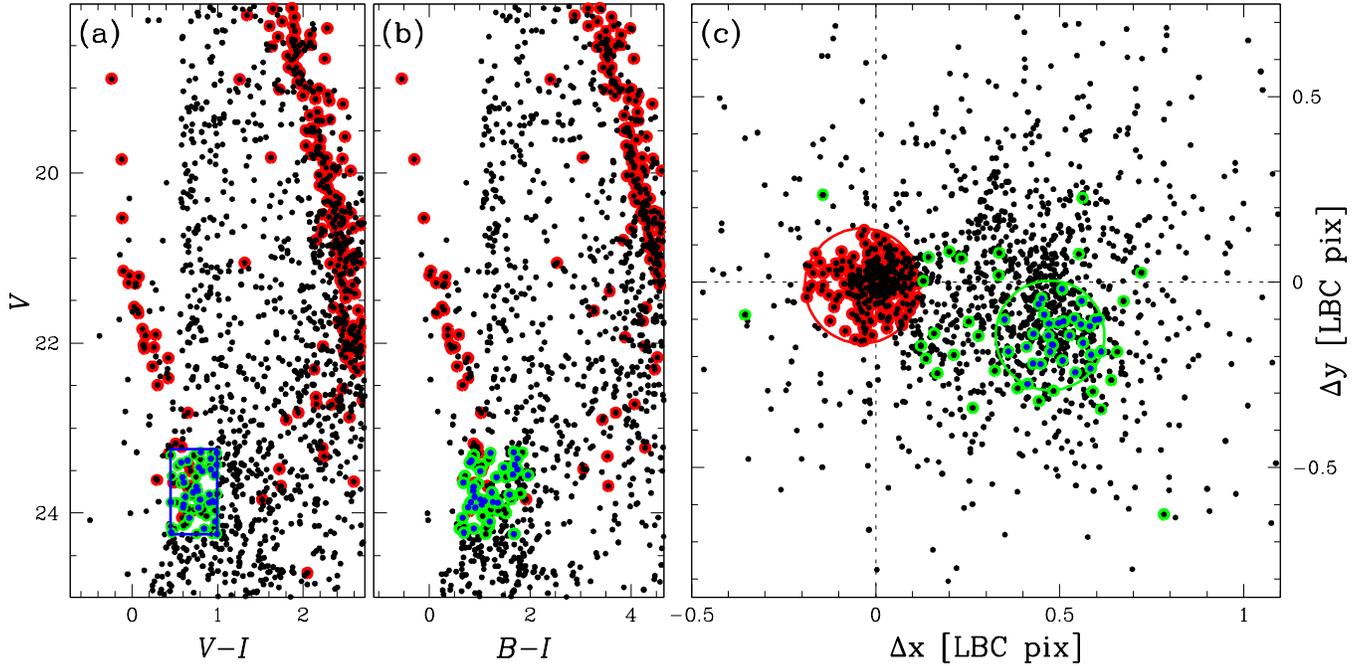}
\caption{\textit{(a)} $V$ vs. $V-I$ CMD zoomed around the end of WD CS
of M~67.  \textit{(b):} $V$ vs. $B-I$ CMD of the same region. In both
panels, black dots are objects with proper motion measurements. Red
circles mark probable M~67 members, as defined in Section
\ref{ss:3.4}.  In (a) we selected a sample of objects (highlighted
with a blue rectangle), superimposed onto the end of the WD CS, which
lie in the region of the CMD used by R98 to define the last two bins
of his LF.  \textit{(c):} show the VPD of all the sources. In red are
M~67 members (within the red circle), and in green color are the
object within the rectangle in panel (a). In blue (in all the three
panels) are those selected objects which proper motion is within 1
$\sigma$ [green circle in panel (c)] around their 3$\sigma$-clipped
median motion.}
\label{f:rich}
\end{figure*}

A thorough study of the WD CS in M~67 was already presented by R98.
In R98, the authors removed background galaxies by mean of
stellar-like shape parameter selections, and by statistically
subtracting objects as measured in a blank field $\sim$$1\fdg2$ away
from the center of M~67. As we can see in our stacked images, the
majority of faint objects are indeed blue compact galaxies, stockpiled
at the same location in the CMD as the end of the WD CS (see Fig.\
\ref{f:cmd_pm}).  We can also see that the faint galaxies are almost
unresolved.  This means that their profiles could mimic stellar
profile, making the shape-like parameter selection criteria not very
efficient.  Moreover, a statistical subtraction of these galaxies
using a blank field is subject to the problem of the cosmic variance
(presence of cosmological structures that alter the statistics of
extra-galactic background objects in different directions, within
limited FoVs).

After the shape selection, R98 extracted the stars for the WD LF
around a 0.7 $M_{\odot}$ CS derived using the interior models of Wood
(\cite{wood95}) and the atmospheres of Bergeron, Wesemael, and
Beauchamp (\cite{bwb95}).  In summary, in R98 WD are never identified.
Their LF has been derived only defining a strip on the CMD for object
selection, and then applying a statistical subtraction of field
objects.  Despite all the aforementioned difficulties, the LF derived
in R98 clearly shows a pile-up around $V$$\simeq$24 [and
$(V-I)_0$$=$0.7], and it terminates at $V$$\simeq$24.25. These values
are in very good agreement with the WD CS termination of
$V$=24.1$\pm$0.1 that we find with our proper-motion selected sample
of WDs.

The sharp peak of the LF identified with the bottom of the WD CS by
R98 needs some more investigation.  We can note, in our $V$ vs. $B-I$
CMD [panel (b) of Fig.\ \ref{f:cmd_pm}], a clump of objects very close to
the end of the WD CS, but slightly redder.  The same objects are
almost superimposed onto the WD CS if we look at the $V$ vs. $V-I$ CMD
used by R98 in their investigation.  The average $V-I$ color of these
objects is $\sim$0.8, close to the $(V-I)_0$$\simeq$0.7, where R98
found a pile-up of objects that were treated by R98 as the bottom of
the WD CS (note that R98 use 0.5 magnitude wide bins in color to
select their WD candidates).  We now further investigate the nature of
these objects.

In Fig.\ \ref{f:rich} we plot our $V$ vs. $V-I$ and $V$ vs. $B-I$ CMDs in
panels (a) and (b), respectively. Black dots are objects for which we
have a proper motion measurement.  Probable M~67 members, as defined
in Section \ref{ss:3.4}, are highlighted in red in both panels.  In
panel (a) we selected for investigation those objects within a (blue)
rectangle, whose borders are defined as $V-I$$=$0.7$\pm$0.25,
$V$=24.25, and $V$=23.25.  This region of the $V$ vs. $V-I$ CMD is the
same one used by R98 to define the last two bins of their WD LF.
Panel (c) displays the VPD of the sources plotted in panels (a) and
(b), on which we have marked in red our proposed M~67 members (stars
within the red circle defined in Section \ref{ss:3.4}), and in green
the objects within the rectangle.  Very few of these objects have a
displacement that is close to the mean M~67 motion, and therefore
could be M~67 stars, but the majority of them is concentrated around a
location in the VPD that is close to the centroid of the motion of
galaxies (Bellini et al. submitted).  Therefore, by means of
proper motion measurements, we conclude that the vast majority of the
objects that form the clump close to the end of the WD CS are
field-type objects, and not M~67 members.

As a further proof that these objects are mainly background galaxies,
we performed the following test.  We calculated the 3$\sigma$-clipped
median displacement (where $\sigma$ is the rms of the residuals around
the median) of these objects on the VPD, and selected the ones within
1 $\sigma$ from this position [big green circle in panel (c)]. This
subsample, constituted by 31 objects, is highlighted in blue in all
the three panels.  We used the LBC $V$ image of 330 s exposure (the
deepest we have, and one with the best seeing $\sim$$0\farcs7$) to
check whether or not these objects are actually galaxies or stars.
Only 21 of them are present in this image. We measured the full width
at half maximum (FWHM) along both LBC $X$ and $Y$ axes.  PSF shapes
and orientation vary with location on the chip, and from chip to chip;
moreover, also galaxy shapes and orientations do vary. Nevertheless,
we expect that, on average, the measured FWHMs of galaxies to be
larger than the ones of stars.

In order to measure FWHMs of stars for this comparison, we selected 10
M~67 WDs in the same $V$ magnitude interval as the investigated
objects. Five of them are present in the 330 s image. We measured the
FWHM for these 5 stars, again along both LBC $X$ and $Y$ axes. We then
measured average values and errors for the FWHMs of stars and
suspected galaxies. The results are as follows:
$$
{\rm stars:~~}
\left\{
\begin{array}{l}
\overline{\rm FWHM}^{\ast}_{X}=3.2\pm0.1\\
\overline{\rm FWHM}^{\ast}_{Y}=2.9\pm0.2\\
\end{array}
\right.
$$
$$
{\rm objects:~~}
\left\{
\begin{array}{l}
\overline{\rm FWHM}^{\rm obj.}_{X}=4.0\pm0.1\\
\overline{\rm FWHM}^{\rm obj.}_{Y}=3.7\pm0.1\\
\end{array}
\right.
$$

It is clear that objects within the green circle in panel (c), are
sizably broader than WDs at the same luminosity, by $\sim$25\% in both
$X$ and $Y$ axes.  On the basis of their average shape and of their
proper motions, we conclude that the objects are indeed blue faint
galaxies and not M~67 stars.  This result does not invalidate what was
done in R98, since we already showed that our estimated WD CS end,
based on a pure sample of M~67 members, is in very good agreement with
the R98 determination.  However, we caution the readers to use the WD
LF to infer information on the M~67 WD properties before an accurate
LF, corrected for completeness, and based on a field-object cleaned WD
CS can be produced.

\section{Comparison with theory}
\label{thWD}

In the same vein as our previous papers on the WD populations in
Galactic star clusters (see, e.g.\ Bedin et al.\ \cite{bedin09} and
references therein), we compare the M~67 WD CS with theoretical WD
models, and also assess the consistency with the results from modeling
of the cluster's MS and turn off (TO) regions.

\begin{figure}[t!]
\centering
\includegraphics[width=\columnwidth]{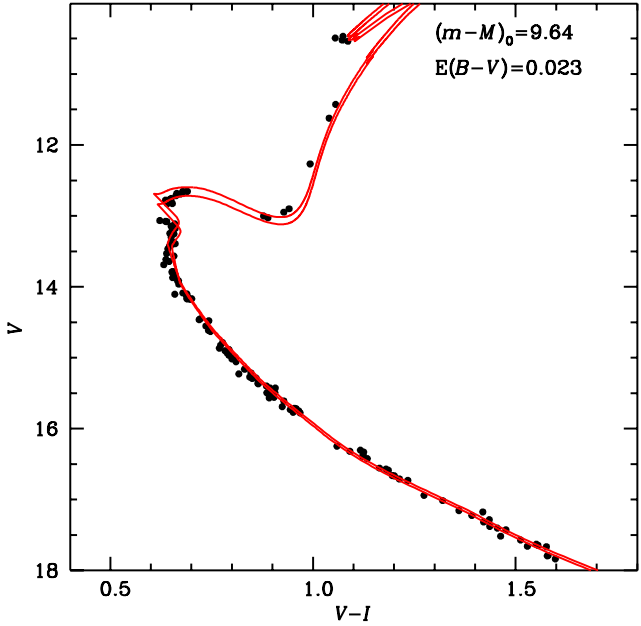} 
\caption{Fit of $V$ vs.\ $(V-I)$ CMD by Sandquist (\cite{sand04}) with
the BaSTI 3.75 and 4 Gyr isochrones with [Fe/H]=+0.06 and convective
core overshooting.}
\label{f:th1}
\end{figure}

As a first step of our analysis, we compare the BaSTI [Fe/H]=+0.06
scaled solar isochrones (Pietrinferni et al.\ \cite{pietr04}) for an
age of 3.75 and 4.00 Gyr to the (single stars only) $V$ vs.\ $V-I$ CMD
by Sandquist (\cite{sand04}) that covers MS, TO, red-giant branch
(RGB) and the central He-burning phases.  As shown in Fig.\ \ref{f:th1},
the best fit to the MS and post-MS phases implies $(m-M)_{0}$=9.64 and
E($B-V$)=0.023 [obtained from the E($V-I$) by using the Cardelli et
al.\ (\cite{cardelli89}) extinction law, with $R_V$=3.1]. These values
are consistent ---within the uncertainties--- with the recent
[Fe/H]=$+$0.05$\pm$0.02 and E($B-V$)=0.04$\pm$0.03 derived by Pancino
et al.\ (\cite{pancino09}), and with the $(m-M)_0$=9.60$\pm$0.09
obtained by Percival \& Salaris (\cite{percival03}).  We will also
adopt these values in the following, for comparisons of the observed
WD sequence with theory.

Had we used a lower [Fe/H] in the fit to the Sandquist data set, the
match to the RGB and red clump would have improved, but at the expense
of the fit to the MS (the slope would not be matched as well as
before). Had we lowered the [Fe/H] value in our isochrones by 0.1 dex
(more than three times the quoted error bars from the spectroscopic
determinations in Pancino et al.\ \cite{pancino09}) the resulting
distance modulus from the overall fit would have increased by only a
few (2--3) hundredths of a magnitude, the reddening by $\lesssim$0.01
mag, and the age would have hardly changed.  All in all, the overall
picture regarding distance, reddening and age would not be changed
appreciably. This works in our favor, because these quantities are
what matters most for the comparison with the WD CS.  Moreover, the
termination of the theoretical WD isochrones would not be really
affected, because the luminosity of the faintest WDs is essentially
driven by their cooling timescales, rather than the progenitor
lifetimes (higher mass progenitors with small life times compared to
the age of the cluster).

Fig.\ \ref{f:th1} shows that the best TO age estimate is 3.8--4.0 Gyr
for BaSTI isochrones with convective core overshooting, in agreement
with previous results (Carraro et al.\ \cite{carraro96} and R98).  A
problem (Carraro et al.\ \cite{carraro96}) persists with the cluster
RGB, which is significantly bluer for the same set of parameters,
independently from the adopted models.  We note that, at the age of
M~67, the mass extension of convective cores in TO stars is small, and
the overshooting extension in the BaSTI models (and in general in all
models including convective core overshooting) is in the regime of
shrinking to zero with decreasing convective core masses (see
Pietrinferni et al.\ \cite{pietr04} for details).

The observed WD sequence is compared to a reference set of
H-atmosphere (DA) CO-core WD isochrones computed using the Salaris et
al.\ (\cite{salaris00}) WD tracks, bolometric corrections from
Bergeron, Wesemael, \& Beauchamp (\cite{bwb95}), progenitor lifetimes
from the BaSTI (Pietrinferni et al.\ \cite{pietr04}) scaled solar
models with [Fe/H]=+0.06 (the same used in the fit to the MS and TO
regions), and the initial-final mass relationship (IFMR) from Salaris
et al.\ (\cite{salaris09}) extrapolated at its lower end to the TO
mass of M~67. We have also employed, as a test, the relationship
proposed by Kalirai et al.\ (\cite{kalirai09}) and found negligible
differences in the resulting isochrones.

\begin{figure}[t!]
\centering
\includegraphics[width=\columnwidth]{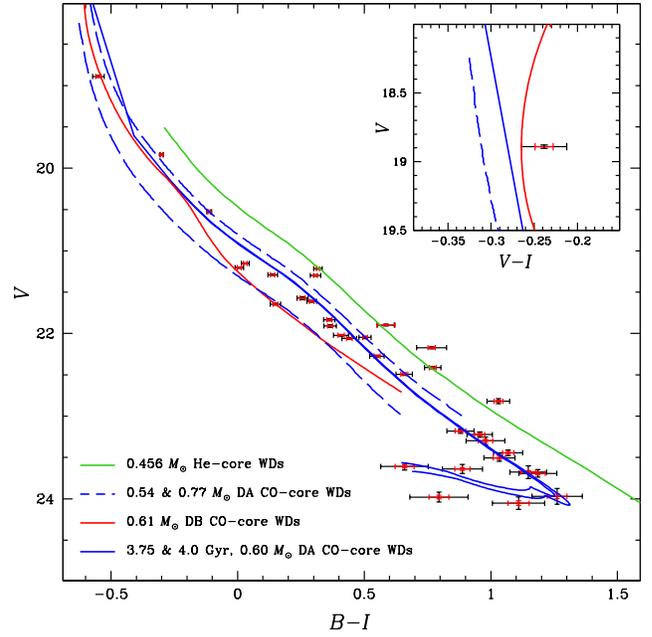}
\caption{$V$ vs.\ $(B-I)$ CMD of the WD sequence compared to our
reference CO-core DA WD isochrones with age of 3.75 and 4 Gyr (blue
solid lines), CO-core DA cooling tracks with masses equal to 0.54
$M_{\odot}$ and 0.77 $M_{\odot}$ (dashed blue lines), a 0.465
$M_{\odot}$ He-core WD (green solid line) and CO-core 0.61 $M_{\odot}$
DB track (red solid line). All isochrones and tracks have been shifted
by the reference distance modulus and reddening values used in Fig.\
\ref{f:th1}. The inset shows the position in the $V$ vs.\ $V-I$ CMD of
the brightest WD in our data compared to the reference isochrones, the
0.77 $M_{\odot}$ DA track and the DB track.  Red error bars are our
formal errors, black error bars are external estimates of the
errors. See text for details.}
\label{f:th2}
\end{figure}

Figure \ref{f:th2} displays our WD isochrones for 3.75 and 4 Gyr, again
using with $(m-M)_{0}$=9.64 and E($B-V$)=0.023 as in Fig.\ \ref{f:th1},
compared to the observed WD cooling sequence in the $V$ vs.\ $B-I$
CMD. The choice of this wide-color baseline enhances differences in
the location of WDs due to variation in their mass or atmosphere
composition, compared to either $B-V$ or $V-I$ ---the other colors
available from our photometry---, and provides more stringent tests
for the theoretical modeling of the observed sequence.  [In addition,
$V$ and\ $B-I$ photometric errors are independent.]  Red error bars in
Fig.\ \ref{f:th2} are the formal photometric errors directly derived
from the uncertainty in the fit slope in Equation~\ref{eq:2} (i.e.,
the internal errors).

Formal errors are generally a lower bound for the true errors.  To
obtain a more reliable ---external--- estimate of the errors, we
randomly divided our photometric data set (for each filter) into four
subsamples.  Specifically, each of the four LBC $B$ subsamples is
constituted by 14 images, each LBC $V$ subsample by 11 images, while
UH8K $I$ images are splitted into 3 groups of 3 images each, and a
last group made with 2 images only (for a total of 11 images).

We selected only the WDs in our sample (see Fig.\ \ref{f:th2}), and for
each subsample separately, we repeated the last reduction step of our
three-step procedures (see Section \ref{ss:3.3}) to compute their
$BVI$ fluxes. For each WD, we obtained four independent measurements
of the flux (in each filter), with formal errors. We weighted the four
estimates of these fluxes according the their formal errors, and took
a weighted mean of the four fluxes.  We found that this mean flux was
equal, within the round-off errors, to the value that we had found
using the whole data set. [It was within few percent in nearly every
case.]  Finally, we derived an external estimate of the true
photometric error, from the residuals of the individual values of the
flux from their mean, using the same weights as we had used for the
mean. These external photometric errors are indicated by the black
error bars in Figure \ref{f:th2}.

\begin{figure}[t!]
\centering
\includegraphics[width=\columnwidth]{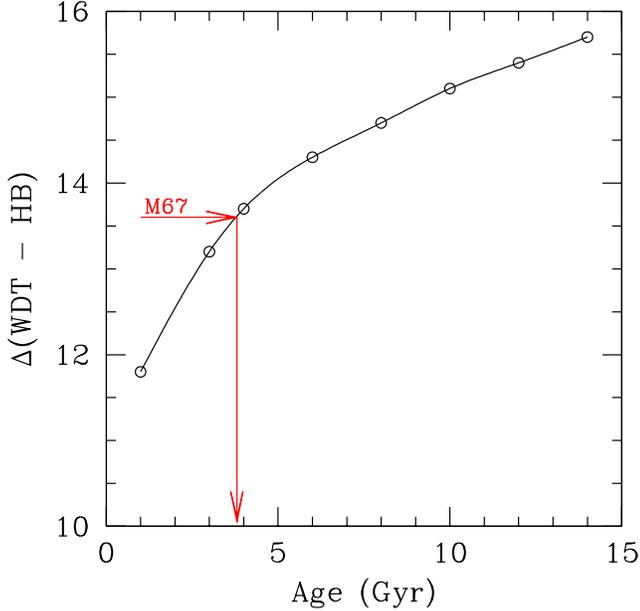} 
\caption{The theoretical value $\Delta V ({\rm WDT - HB})$ is plotted
against the age of the simulated cluster. The value observed for M~67
in the present work is also shown.}
\label{f:th3}
\end{figure}

We note, instead, that error estimates based on artificial stars are
also (as our formal internal errors) generally underestimated, and so
less reliable.  This is because artificial stars are always based on
an input PSF model, which could significantly differ from the true
local PSF, concealing our ignorance about the true sources' flux.
Moreover, thanks to the large dither pattern of LBC observations, WDs
fall each time in different locations on the LBC chips, in different
chips as well, and under different observing conditions, contributing
to further strengthening the reliability of the external estimates of
the errors.

There is good agreement between the location of the WD isochrones and
the observed sequence. The expected blue hook at the bottom of the WD
sequence is also visible in the data. It is due to the presence of
increasingly massive (lower radius) objects originating from more
massive MS progenitors that had more time to cool down along the WD
sequence. The location and color extension of the blue hook is well
matched by the isochrones with ages consistent with the TO age. We
recall that the observed $V$ magnitude of the bottom end of the WD
sequence ($V$$\sim$ 24.1$\pm$0.1 mag) is in very good agreement with
the observed WD LF by R98, and with the predictions by Brocato,
Castellani \& Romaniello (\cite{bcr99}, their Fig.\ 8), where an age
of 4 Gyr, solar metallicity and a distance modulus $(m-M)_{0}$=9.6
were assumed.

A complementary method to estimate the age of M~67 uses a new age
indicator, that has the advantage of being free from distance and
extinction effects, and it is also independent from the age obtained
from the TO.  Similarly to the $\Delta V$ parameter employed in
globular cluster studies ($V$-magnitude difference between the
Horizontal Branch and Turn Off, see e.g., the review by Stetson,
Vandenberg \& Bolte \cite{stetson96}) it is possible to define the
parameter $\Delta V ({\rm WDT - HB})$ as the difference between the
$V$ magnitude of the WD CS end [$V$(WDT)], and the mean level of the
horizontal branch [$V$(HB) ---see Brocato, Castellani \& Romaniello
\cite{bcr99}].  Once the metallicity of the cluster is known, this
parameter is a function of the cluster age, increasing for older
clusters (see Fig.\ \ref{f:th3}).  Here, we present the calibration of
$\Delta V ({\rm WDT - HB})$ as a function of age for M~67 metallicity,
while the general calibration for a wider range of chemical
compositions will be presented in a forthcoming paper (Brocato et
al. \ in prep.).  Here it suffices to say that our data provide
$\Delta V ({\rm WDT - HB})$=13.6$\pm$0.1, that leads to an age of
3.9$\pm$0.1 Gyr. This estimate (which is formally independent of the
CMD fitting procedure used in Fig.\ \ref{f:th1} to estimate the cluster
age) is in agreement with the other estimates above discussed.

\section{Discussion}

Despite the good agreement between the location of the WD isochrone
and the observed CS, and the consistency between TO and WD ages, in
the $V$ magnitude interval between $\sim$19 and $\sim$23 there are
several objects that appear more than 3$\sigma$ away from the
theoretical sequence, and deserve a further analysis.  We cannot
exclude, a-priori, that these objects are WD+WD binaries, but this
possibility is very unlikely, therefore in the following we will not
treat this case.  We have first considered the possible presence of a
large spread in the cluster IFMR (see, e.g., Salaris et al.\
\cite{salaris09} and Dobbie et al.\ \cite{dobbie09}) for progenitor
masses in the range $\sim$1.5--2.5 $M_{\odot}$.  Figure \ref{f:th2}
shows the cooling tracks of 0.54 $M_{\odot}$ and 0.77 $M_{\odot}$ DA
CO-core WDs (dashed blue lines).  In the appropriate magnitude range
these two models are located on the red and blue side of the reference
isochrone, respectively, that is populated by WDs with mass $\sim$0.6
$M_{\odot}$. A spread of the IFMR towards higher values of WD masses
can be explained by the spread of objects on the blue side of the
isochrone, whereas the redder WDs are not so easy to explain. The 0.54
$M_{\odot}$ track is still about 0.1 mag bluer than 6 objects with $V$
between $\sim$19 and $\sim$23. Taking into account that the core mass
at the first thermal pulse is $\sim$0.52--0.53 $M_{\odot}$ for the
relevant progenitor mass range, it is difficult to match the position
of these red objects with ``standard'' WD sequences. A possible way
out of this problem is to invoke the presence of CO-core WDs with
masses lower than the core masses at the beginning of the thermal
pulse phase. As studied by Prada Moroni \& Straniero (\cite{prada09}),
anomalous and somewhat finely tuned mass loss processes during the RGB
phase can produce CO-core WDs with masses as small as $\sim$0.35
$M_{\odot}$ at the metallicity of M~67, for progenitors around 2--2.3
$M_{\odot}$. In this mass range there is a transition between electron
degenerate and non degenerate He-cores along the RGB. These low mass
WDs are potentially able to match the handful of objects redder than
our reference isochrone. Their cooling timescales (e.g.\ Fig.\ 12 in
Prada Moroni \& Straniero \cite{prada09}) are roughly consistent with
their observed luminosities, assuming that they are produced by
progenitors with masses around 2--2.3 $M_{\odot}$.

There is another possible interpretation.  In order to reproduce the
position of these WDs in the CMD, we have considered the cooling track
(green line in Fig.\ \ref{f:th2}) of the 0.465 $M_{\odot}$ He-core WD
discussed in Bedin et al.\ (\cite{bedin08a}).  At the cluster's
metallicity, BaSTI models predict that the He-core mass of stars
climbing the RGB (mass $\sim$1.4 $M_{\odot}$) can be at most equal to
$\sim$0.47 $M_{\odot}$. This WD track represents therefore an
approximate bluer boundary for the location of M~67 He-core WDs. All
of the 6 red objects (that are more than 3$\sigma$ away from the
CO-core WD isochrone) lie close to this sequence.  They can be massive
He-core WDs produced by very efficient mass loss along the RGB
phase. These same 6 objects are located around the He-core sequence
also in $V-I$ colors although the separation between the sequences is
smaller in that case (and even smaller in $B-V$).  As for the objects
at the blue side of the isochrones, Fig.\ \ref{f:th2} displays the
cooling track of a 0.61 $M_{\odot}$ CO-core model with pure
He-atmospheres (DB, red solid line). This is approximately the
expected value of the DB mass evolving down to $V$$\sim$23 (where the
track is truncated) for the chosen progenitor lifetimes and IFMR.  At
fainter magnitudes, DA and DB tracks of the same mass have
approximately the same colors.

There is clearly one object at $V$$\sim$19, and two objects with $V$
between $\sim$21 and 22, that are well matched by the DB model and are
therefore candidates to be He-atmosphere WDs.  It is clear that
without spectroscopic estimates of atmospheric composition and/or WD
masses it is impossible to find a unique interpretation for these WDs
that are not matched by the reference isochrone. But both $B-I$ and
$V-I$ colors suggest that the brightest object at $V$$\sim$19 is a DB
WD. The inset of Fig.\ \ref{f:th2} displays the position of this star in
the $V$ vs.\ $V-I$ CMD. In this CMD the bright part of the DB sequence
is located on the red side of the reference DA isochrone, whereas at
magnitudes fainter than $V$$\sim$20 the relative position of the two
sequences is the same as in the $B-I$. The star is located at the red
side of the DA isochrone in $V-I$, and at the blue side of the
isochrone in the $B-I$, exactly as expected for DB objects, and in
complete disagreement with its interpretation as a massive DA object.

\begin{acknowledgements}

A.\ Bellini\ acknowledges support by the CA.RI.PA.RO.\ foundation, and
by the STScI under the \textit{``2008 graduate research
assistantship''} program.  G.\ P.\ acknowledges partial support by
PRIN07 (prot. 20075TP5K9).

\end{acknowledgements}

{}

\end{document}